# Unified Architecture for Data-Driven Metadata Tagging of Building Automation Systems

Sakshi Mishra[1*], Andrew Glaws[2], Dylan Cutler[3], Stephen Frank[4], Muhammad Azam[5], Farzam Mohammadi[6], Jean-Simon Venne[7]

[1,2,3,4]National Renewable Energy Laboratory, Golden, USA

[5,6,7]BrainBoxAi, Montreal, Canada

Emails: [1]Sakshi.Mishra@nrel.gov, [2]Andrew.Glaws@nrel.gov, [3]Dylan.Cutler@nrel.gov, [4]Stephen.Frank@nrel.gov, [5]m.azam@brainboxai.com, [6]f.mohammadi@brainboxai.com, [7]js.venne@brainboxai.com

*corresponding author

***Abstract***— This article presents a Unified Architecture (UA) for automated point tagging of Building Automation System (BAS) data, based on a combination of data-driven approaches. Advanced energy analytics applications—including fault detection and diagnostics and supervisory control—have emerged as a significant opportunity for improving the performance of our built environment. Effective application of these analytics depends on harnessing structured data from the various building control and monitoring systems, but typical BAS implementations do not employ any standardized metadata schema. While standards such as Project Haystack and Brick Schema have been developed to address this issue, the process of structuring the data, i.e., tagging the points to apply a standard metadata schema, has, to date, been a manual process. This process is typically costly, labor-intensive, and error-prone. In this work we address this gap by proposing a UA that automates the process of point tagging by leveraging the data accessible through connection to the BAS, including time-series data and the raw point names. The UA intertwines supervised classification and unsupervised clustering techniques from machine learning and leverages both their deterministic and probabilistic outputs to inform the point tagging process. Furthermore, we extend the UA to embed additional input and output data-processing modules that are designed to address the challenges associated with the real-time deployment of this automation solution. We test the UA on two datasets for real-life buildings: (i) commercial retail buildings and (ii) office buildings from the National Renewable Energy Laboratory (NREL) campus. The proposed methodology correctly applied 85-90% and 70-75% of the tags in each of these test scenarios, respectively for two significantly different building types used for testing UA's fully-functional prototype. The proposed UA, therefore, offers promising approach for automatically tagging BAS data as it reaches close to 90% accuracy. Further building upon this framework to algorithmically identify the equipment type and their relationships is an apt future research direction to pursue.

1. INTRODUCTION

   *1.1. Motivation*

Building management systems, also known as Building Automation Systems (BAS) are a combination of hardware and software layers forming a fully functional control system for monitoring and controlling a building's electrical and mechanical equipment. A BAS consists of various sensor points, setpoints, and command points on distributed controllers that coordinate with higher-level controllers using protocols such as BACnet [1], LonWorks [2], Niagara (Fox protocol) [3], or KNX [4]. The BAS may optionally include upper-level software that provides a programming interface for control sequence development, a data historian, and/or a human-machine interface. In addition to a typical BAS, other peripheral monitoring and control systems that collect relevant operational data, such as electrical metering systems, are often present in a building. These auxiliary systems may communicate with the BAS via protocol translation or hardware gateways, or may provide independent data collection pathways. BAS are a subset of Energy Management Information Systems (EMIS), which are defined as "combined hardware and software products that comprise a broad family of tools and services to manage commercial building energy use" [5]. Systems such as BAS, Fault Detection and Diagnostics, and Automated System Optimization fall under the broad umbrella of EMIS.

BAS are standard for large commercial buildings and offer a myriad of benefits. From a building manager perspective, programmatic control of building systems, remote monitoring of the equipment (such as electrical supply and air handling units (AHUs)), and associated ease of maintenance are among the major benefits. From an occupant perspective, the benefits include increased comfort. From a building owner's perspective, energy cost reduction is the most promising benefit. Additionally, BAS vendors often include a supervisory interface that may offer features such as data consolidation, report generation, fault detection, or predictive maintenance. However, if present, these features are typically vendor-specific and historically have rarely followed any common standard for data organization.

BAS specifically, and EMIS more broadly, are key to Grid-interactive Efficient Buildings (GEB) [6]. When a critical mass of urban built structures converts to GEB, they have the potential to alter the profile of power system demand in major load pockets. This flexibility can lead to lower cost or lower carbon grid operations and bolster large-scale renewable energy integration. However, the lack of informational interoperability remains a key barrier that prevents disparate systems from working together seamlessly to achieve GEB goals [7, 8]. Metadata tagging based on a standardized schema provides a possible path to overcoming these interoperability issues. The challenge in this approach lies in the tedious and expensive process of manually tagging all of the buildings sector.

The U.S. commercial buildings sector has buildings with BAS installed covering about 42% of the total floor area of this sector [9]. Given the current and forecasted adoption levels for BAS and their foundational role in enabling a GEB vision for our built environment it is very important to address the issue of lack of informational operability between buildings (and at times, different parts of the same building) due to lack of common tagging schema. Automatically applying a standardized metadata schema is a crucial need to accelerate the effective application of analytics and enable buildings' participation in the GEB future. The data which comes from BAS needs to be interpreted systematically by machines to enable effective analytics, external supervisory control, and other internet of things (IoT) applications. Historically, any given building's BAS point names follow the in-house convention applied by the controls contractor who installed the system, rather than any set standard. The only metadata applied to the BAS points is a "units" category. Another piece of informative data is the BACnet Object Type (Analog Input, etc.).

Within the past decade, the informational metadata standards Project Haystack [10] and Brick Schema [11] (informally, "Haystack" and "Brick") have emerged and matured. These standards provide structured semantic metadata for building systems, equipment, and points. Assignment of semantic tags per the standards greatly reduces the implementation barriers for advanced control and analytics applications, including FDD, supervisory control, and GEB applications. However, the implementation of Haystack or Brick for existing BAS and EMIS requires a mapping process by which each data entity is assigned descriptive tags. Traditionally, this tagging process has been executed manually by an engineer. It is a time-consuming, labor-intensive, and error-prone process. According to Granderson et al. [12], the software cost of integrating and maintaining Energy Information System (EIS) systems can range from $230/point (upfront) to $1880/point (5-year ownership). Given such high cost associated with tagging and maintaining the BAS data manually, it also acts as one of the barriers for automated fault detection and diagnosis technology for small commercial buildings [13]. Automating this process of applying tags to the BAS points to generate metadata associated them has the potential to expedite the timeline of tagging from *months* after commissioning to *weeks* or in some cases *days*.

*1.2. Literature Review*

Over the past few years, there has been growing interest in the automated application of metadata to BAS objects to efficiently enable the necessary interoperability to achieve GEB goals. Early approaches focused on raw text strings applied to devices during installation. These point names tend to be highly variable among buildings and are often character-limited such that the amount of information encoded into these names is restricted. Bhattacharya et al. [14] propose a method for generating rules to parse these raw point names and reformulate them as normalized labels with the aid of expert intervention. A syntactic clustering identifies highly informative points to query the expert to minimize the level of human intervention. Because many tags within a given building tend to follow similar syntactic formats (either because installed points come from the same manufacturer or were labeled by the same contractor), this approach can effectively identify a majority of these points with relatively few examples. However, points with atypical naming conventions remain elusive. Furthermore, this approach requires syntactic consistency in point names across buildings in order to be transferred to new facilities without repeating the learning step. This approach has three major caveats - i) human participation is a necessary part of the process; ii) it is not effective for the points with atypical naming conventions, and iii) learning step is necessary if the syntactical consistency doesn't exist within different buildings.

Schumann et al. [15] construct a similarity metric for comparing raw point names to entries in a manually constructed dictionary. This produces ordered lists of point labels based on the highest similarity scores that can then be more quickly processed by a human expert. This approach is heavily reliant on the quality of the constructed dictionary and struggles when it encounters naming conventions that are not included in the dictionary. Hong et al. [16] propose an active learning approach that clusters raw point names within a given building, then queries the user for labels that will be most informative in terms of automatically propagating labels to nearby points. This procedure assumes the similarity of the point names within each building. Firstly, given that the larger part of the processes for the above two approaches is conducted through the human participation, their error proneness increases. Moreover, due to reliance on the quality of the constructed dictionary, it clearly doesn't offer a solution that can be deployed and scaled in real-life multiple buildings settings.

In addition to raw text, many points also generate data streams that describe the point's real-time measurements or state. These data streams offer a promising path to applying point metadata because their characteristics tend to be more consistent across buildings. For example, Figure 1 shows unlabeled time-series data for several points from the buildings considered in this work. The time-series data generated from these points exhibit distinct behaviors. These behaviors can be used to infer semantic facts about the points. For example, point A cycles between 1 and 0, suggesting a binary sensor or control signal. Point B varies continuously, suggesting a sensor. Point C varies discretely with a daily cycle. This behavior and the magnitude range suggest that Point C may be a temperature setpoint.

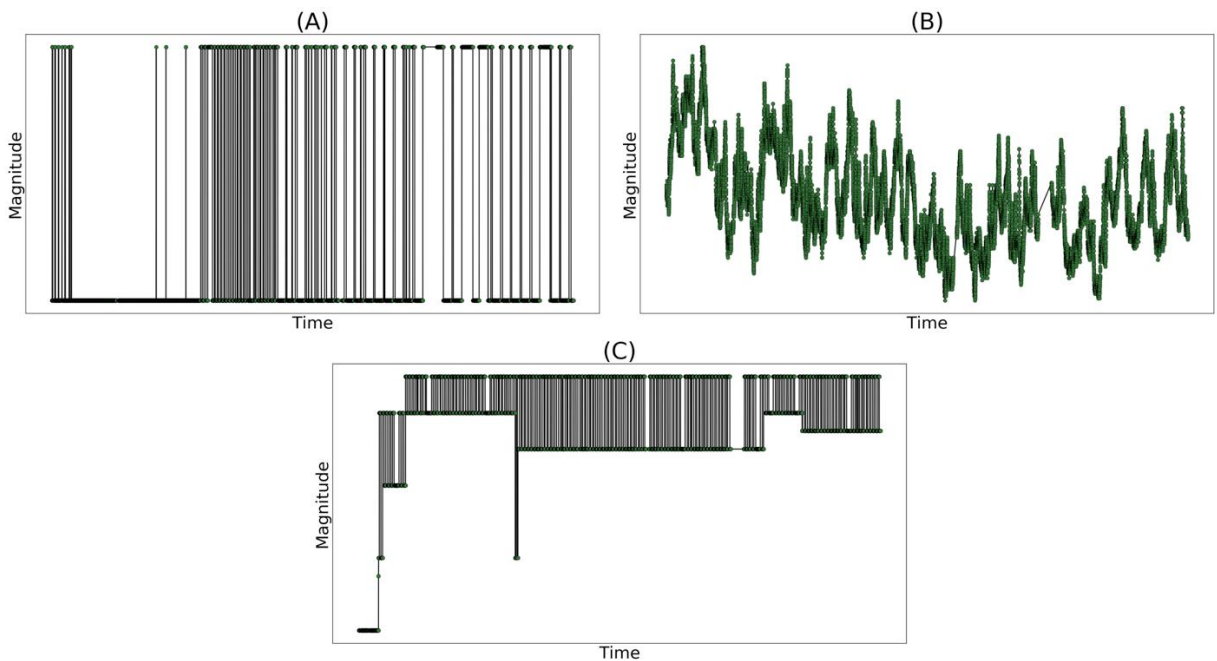

Figure 1 – Example time-series data for several points

Automatically analyzing time-series data for metadata generation requires the identification and quantification of the unique characteristics of each data stream. In general, machine-learning-based approaches to time-series data analysis can be divided into two steps: (i) characterization of descriptive data features and (ii) clustering or classification of the data based on these features [17]. Methods for building metadata generation based on time-series analysis have explored various techniques for performing these two tasks, as summarized in the following paragraphs.

Early work analyzing these data streams focused on how to infer the physical locations of the various devices within a building. Such information can be useful for mapping relationships between points. Hong et al. [18] attempt to use a frequency analysis of the data streams to infer whether or not pairs of sensors are located in the same room. The data are transformed using empirical mode decomposition (EMD) into intrinsic mode functions (IMFs), which characterize various frequencies in the data [19]. By examining the correlation coefficient of IMFs with appropriate frequencies, Hong et al. show that statistical boundaries exist in the data that are predictive of physical boundaries in the building. Koc et al. [20] examine the effects of data size and whether a non-linear correlation coefficient is more effective at determining sensor locations. Akinci et al. [21] identify the specific rooms in which sensors are located by correlating expected and actual sensor readings with the known HVAC energy input into a room. By identifying a single sensor in a room, this process can be identified with previous work to map all sensor locations in a building. However, this approach requires some a priori knowledge regarding the building's HVAC system, such as the building layout and the mapping of heating/cooling setpoints – calling for increased human intervention.

While the physical location is a key component of understanding the interactivity between devices in a building, fully descriptive metadata encompasses a wider range of features, which the approaches mentioned in the previous paragraphs do not capture. Automated methods for generating these data must be able to identify the type of point, what physical quantity it is measuring or actuator it is controlling, and what other devices it is connected with, in addition to its physical location. Recall that time-series analysis depends both on the methods for characterizing the data and for classifying or clustering those characterizations. For much of the work done in this field, the classification or clustering techniques are based on classical machine learning algorithms. The approaches for characterizing the data stream tend to be more varied and generally leverage three main toolsets: (i) descriptive statistics, (ii) derivative-based methods, and (iii) frequency analysis. Though these approaches were successful to some degree, in finding the spatial context/physical location of the devices, they do not offer a way to generate required extensive metadata associated with each point in the BAS system.

Gao et al. [22] compare several standard machine learning classification techniques such as random forests, support vector machines, and naïve Bayes for applying Haystack tags. The feature vectors for the machine learning methods are simply the mean, median, mode, quantiles, and deciles of each data stream. This work concludes that the random forest is the most robust algorithm for applying these tags. Additionally, the authors compare performance when applying individual tags versus composite tags (analogous to "tagsets" in Brick) and conclude that individual tags can be more accurately applied. This is because using composite tags causes the output space to explode while using individual tags increases the amount of information associated with each tag given that multiple point types may share common individual tags. The contribution of this work is noteworthy in its attempt to explore the application of modern machine learning classification techniques and comparing the performance of various choices available with this class of algorithms. However, this work doesn't present a way to harness the synergies offered by the orchestrated application of these algorithms. Overall, the proposed solution, in its current state, is not a research prototype that addresses the challenge in its entirety.

Calbimonte et al. [23] characterize the time-series data using local linear approximations to estimate derivatives. The local linear models are constructed adaptively using a greedy algorithm to minimize the number of piecewise linear segments used to approximate the data. The distributions of derivatives are used to compare the various points to each other using a k-nearest neighbor (kNN) algorithm. Semantic representations of metadata of known points can then be propagated to their nearest neighbors. Holmeggard and Kjærgaard [24] compare slopes and EMDs of data streams by computing the cosine of the angle between these vectorizations. This cosine distance is combined with another measure called dynamic time warping (DTW) that accounts for differences in speed between two sequences of times series data. Metadata is then propagated from the nearest known point.

More recently, efforts have been made to effectively combine information from the raw point name and the time-series data in order to improve performance; plaster [25] and scrabble [26] are a few notable examples. Bhattacharya et al. [27] expand on the work of [14] by using statistical characteristics (i.e., median and variance) of the time-series data to identify points that are likely to be similar despite differences in the raw name data. Expert labels for known points can then be extended to these candidate points to cover more outliers. Balaji et al. [28] and Hong et al. [29] both propose transfer learning approaches that cluster raw point name data to simplify the metadata generation process. In the former, the time-series features are constructed for each cluster using statistical and frequency-based techniques. In the latter, the unlabeled clusters inform an ensemble of classifiers that analyze the statistical characteristics of the time-series data in order to generate metadata. Some commercial entities are applying this approach in real buildings [30]. This hybrid line-of-inquiry, utilizing both raw point names and time-series data is an interesting step toward filling the long-existing void. Yet, these approaches leave room for improvement to research a methodology that leverages different types of machine learning algorithms (supervised and unsupervised) to effectively harness the information stored in these two streams of data.

It can be inferred from the literature works discussed in the above paragraphs that the previously published works have tackled the problem of tagging the BAS points automatically with a piece-meal approach, that is, they evaluated the effectiveness of an individual algorithm or approach in automatically tagging the points. In other words, these works are focused on applying one type of algorithm (either rule-based or statistical) to one type of building data (e.g., university campus) data only. Moreover, there seems to be a clear chasm between two lines of inquires in the large percentage of works in the literature – one based on raw point names and other time-series data. This exclusivity prevents these approaches from thoroughly harnessing the information at hand (point name and their associated time-series data) for making completely automated metadata tagging decisions. Although they have had a varying degree of success in addressing the challenge under study, none of them have addressed the problem from a "real-life implementation" perspective. The question of building an integrated workflow that can harness the strengths and complement the weaknesses of various algorithms, by leveraging both point name and time-series data, to effectively tag the BAS data remains relatively unexplored.

Moreover, not many works have focused on exploring challenges associated with building a software architecture that can be generalized over more than one type of building. Therefore, in this work we take a holistic look at the problem of tagging metadata automatically, focusing on real-time deployment challenges, by attempting to answer the questions such as: What kind and amount of data is typically available from the BAS before the metadata tagging is performed? In other words, is it a new building or has it been operational for a few months? What are the similarities and differences between the data from different types of buildings? How can both the raw point names and the time-series data be analyzed in parallel to arrive at the decision of assigning a tag? What is the scalability of the proposed solution?

*1.3. Manuscript Structure and Contributions*

We propose a Unified Architecture (UA) that focuses on solving the problem of automated BAS point metadata tagging. Within the overarching research topic of automated building metadata tagging, we:
- Propose a holistic framework for generating metadata tags for BAS points that describes three distinct phases of metadata generation;
- Identify challenges associated with real-life application of automated tagging solutions on various types of buildings and different use-cases;

- Propose a UA for point tag identification informed by a set of supervised and unsupervised machine learning algorithms, capable of embedding human expert knowledge in the framework using rule-based blocks; and
- Present a detailed data-postprocessing framework to enhance the effectiveness of the proposed UA over time, from a real-time deployment perspective.

The organization of the remainder of this article is as follows: Section 2.1 provides an overview of the automated point tagging problem. It begins with a discussion of the various tagging schemas that have been proposed in the literature and describes the Haystack tagging standard that is applied in this UA approach, and concludes by defining the multiple aspects of the automated point tagging problem. Section 3 highlights the data considerations for this problem, including data collection methods, and discusses how data availability and building type impact the choice of approach for automatically tagging the BAS. It leads to a detailed discussion provided in the Appendix about identifying the challenges associated with deploying an automated solution in the real-world where building types and data availability varies drastically. An overview of various machine learning algorithms (supervised and unsupervised) and rule-based workflows that are employed in this research, is provided in Section 4. Section 5 presents the proposed United Architecture and explains its components in detail. The results of the two case studies conducted using the proposed UA are presented in Section 6, along with a discussion around the existing limitations of the proposed UA. Conclusions and future research directions are described in Section 7.

2. BAS TAGGING AND PROBLEM DEFINITION

*2.1. BAS Tagging Schemes – an overview*

As advanced analytics and controls applications become more prevalent, the need for a cross-cutting industry standard for organizing point metadata is becoming increasingly important. BAS schemas are constantly being developed or reconfigured [31]. Because any automated approach to generating BAS metadata will be schema-specific, it is important to understand the key features of the available schema. Here, we provide an overview of several BAS schemas, specifically BASont, SAREF, Brick Schema, and Haystack.[1]

BASont is a building ontology model built on the Industry Foundation Classes (IFC) standard for describing device instances in buildings [32]. IFC provides a framework for interoperability among building architecture, construction, and management software [33]. BASont builds high-level templates of rooms and equipment that broadly describe the structure and functionality of devices while point-specific details are applied at lower levels to maintain flexibility. This template-based approach enables scaling to large commercial buildings. Additionally, device replacement is simplified since unchanged information is quickly applied to the new point via templates while changed information (e.g., manufacturer-related information) can be updated individually.

The Smart Appliances Reference (SAREF) ontology, developed by the European Commission (EC) and the European Telecommunications Standards Institute (ETSI), captures the functional relationships between smart devices [34]. This ontology considers devices (the physical objects in a building) and the possible functions they can perform. SAREF defines a broad array of simple functions that act as building blocks for constructing more specialized and complex functions. A device's function is expressed through the service the device provides and changes its state when called (e.g., a light switch can turn on or off). Additionally, devices contain the knowledge of their energy usage in various states to enable intelligent decision-making for energy efficiency.

Brick Schema is a newer and more robust metadata schema that provides an ontology to describe HVAC, lighting, and power infrastructure in commercial buildings [35, 36]. Brick leverages the structure of the Resource Description Framework (RDF) [37] specification to define a standard organization schema for these data. The schema uses triples of descriptive tags with the subject-predicate-object format standard to RDF that can be applied to various entities in a building to describe location, functionality, and relationships to other entities. In general, Brick triples describe one of four key facets of any device:

1. *Point* – the type of physical or virtual entity that is generating data related to some facet of the physical space,
2. *Equipment* – larger devices that communicate with multiple points to accomplish some task within the building,
3. *Location* – the physical space in the building where a point is located,
4. *Resource* – the physical material that interacts with the point.

---

[1] We note that Project Haystack, Brick Schema, and members of the ASHRAE 223P standards committee are collaborating to define a formal ontology, or a family of formal ontologies, for BAS metadata. The goal is to release ASHRAE Standard 223P, *Designation and Classification of Semantic Tags for Building Data*, as an ISO standard. We anticipate that this collaboration will produce improvement of and alignment in the Haystack and Brick standards. However, in this article we address the current, published versions of these standards only. Because at the time of this writing the ASHRAE 223P committee had not yet released a draft, we also omit discussion of ASHRAE Standard 223P.

Brick organizes entities into fixed classes, which have expected tags and relationships associated with them. This framework maintains a relatively simple structure while allowing for significant flexibility in describing points within a building. Furthermore, tags can inherit properties of other higher-level tags and can be combined to create tagsets (e.g., a point can be tagged as a `zone temperature sensor`), which are often associated with specific entity classes. Relationships in Brick attempt to capture connections or encapsulations. The central relationships (predicates) are defined by *isLocationOf*, *controls*, *hasPart*, *hasPoint*, and *feeds* (as well as their inverse relationships). The tag and relationship model of Brick makes it simple to understand and allows it to cover a wide range of possible use-cases. However, Brick does not recommend ad-hoc tags or relationships, which makes extending the standard to cover new technology areas more difficult. Brick also has limited commercial (non-research) adoption.

Haystack [10] is an open-source initiative first developed in 2014 through the collaboration of multiple industry partners seeking to construct a robust but flexible semantic model to building metadata. Although Haystack encodes similar information as Brick schema, it does not enforce a formal class system like Brick does. Rather, Haystack employs collections of tags to describe entities and the relationships between them. Each tag is associated with (assigned to) an entity. Haystack permits the ad-hoc addition of non-standard tags, which makes it readily extensible in practical applications.

Tags in Haystack are constructed with name-value (or key-value) pairs. Haystack supports thirteen atomic types (or "kinds") of tags. Although all atomic kinds are technically distinct, conceptually they can be organized loosely into three categories: (i) marker tags, or name-only singletons, that indicate intrinsic properties of an entity (i.e., entity *type* or *is-a* descriptions); (ii) value tags that describe entity properties with associated values; and (iii) reference tags that describe relationships between entities. The "Marker" and "Ref" kinds formally define markers and references, while all other kinds represent the broader category of value tags. The value in the name-value pair of a Marker tag is a generic annotation with no associated meaning; the value of a Ref tag is the unique identifier of an entity. The special Ref tag `id` defines an entity's unique identifier; this `id` tag becomes the target for Ref tags on other related entities.[2]

Most of the semantic meaning in Haystack is described via markers and value tags. For example, a given point may have the `zone`, `temp`, and `sensor` markers applied to it, indicating that it is a sensor that measures a zone temperature. The same tag may have the Str (string) tag `unit: "°C"` applied to let us know how the sensor is measuring temperature. Tag values can be strings, numbers, Booleans, dates, or times, among other data types. For relationships, Haystack employs the basic hierarchical structure `site`, `equip`, `point`: a `site` contains multiple `equip`s (equipment), and an `equip` contains multiple `point`s. These key relationships are described using the `siteRef` and `equipRef` tags. Other Ref tags specify other relationships, such as between a main meter and its submeters, between an AHU and its associated variable air volume (VAV) boxes, etc.

At present, the Haystack standard defines mutually exclusive sets of tags in human-readable documentation, but not in a machine-readable format. Therefore, an algorithm implementor must translate these tag sets into machine-readable rules. With respect to this aspect of implementation, the formal structure of Brick schema offers the advantage of machine-readable tag sets with a formal class hierarchy, or ontology. The advantage of a formal ontology is that it allows greater exploitation of the structure of the data model and the associated human knowledge embedded within it; the disadvantage is that it makes it more difficult to extend the standard with the addition of ad-hoc tags. Versions 4+ of the Haystack standard are expected to include an optional ontology with machine-readable documentation for tag sets that will allow automated synthesis of tagging rules, however, at the time of this writing Haystack version 4 had not been released.

However, the Haystack tagging schema has several key benefits that make it an appropriate choice for the automated metadata generation algorithm presented here. First, the generalized tagging schema makes Haystack a very simple yet flexible model for building metadata. The tag-based approach is effectively machine- and human-readable so that it can integrate easily with standard machine learning tools and produce interpretable results. This is in contrast to the more rigid ontology-based schemas such as BASont and SAREF. Additionally, Haystack tags are meant to define an extensible schematic model that is scalable to large buildings containing hundreds or thousands of devices. Users can create their own ad-hoc tags to address unique scenarios arising in any building environment without forcing a modification to the underlying semantic model. To accommodate this flexibility in the Haystack model, the metadata algorithm should be flexible to deal with unique tagging situations as needed. Finally, and most importantly for practical use, Haystack has strong baseline popularity compared with competing standards. Haystack has emerged as a common choice for teams and companies implementing analytics and other supervisory software on top of BAS and many software platforms and products now support Haystack. For these reasons, we selected Haystack as the basis for the present work. However, we also note that Haystack's flexibility and lack of formal class system present some specific challenges that we discuss in Section 4.1.

---

[2] Throughout the manuscript, we use monospace font to indicate a Haystack tag, as in: `sensor`.

*2.2 Problem Definition*

The overarching goal of the tagging process is to identify all of the data acquisition points contextually to enable further development of analytics layers on top of the BAS. Conceptually, the tagging process for the Haystack tagging schema consists of three general steps (Figure 2):

- Point identification: Classify the type or characteristics of each individual point.
- Equipment identification: Group associated points into equipment entities and identify the type or characteristics of the equipment.
- System identification: Identify and classify relationships among equipment and map these relationships into building systems.

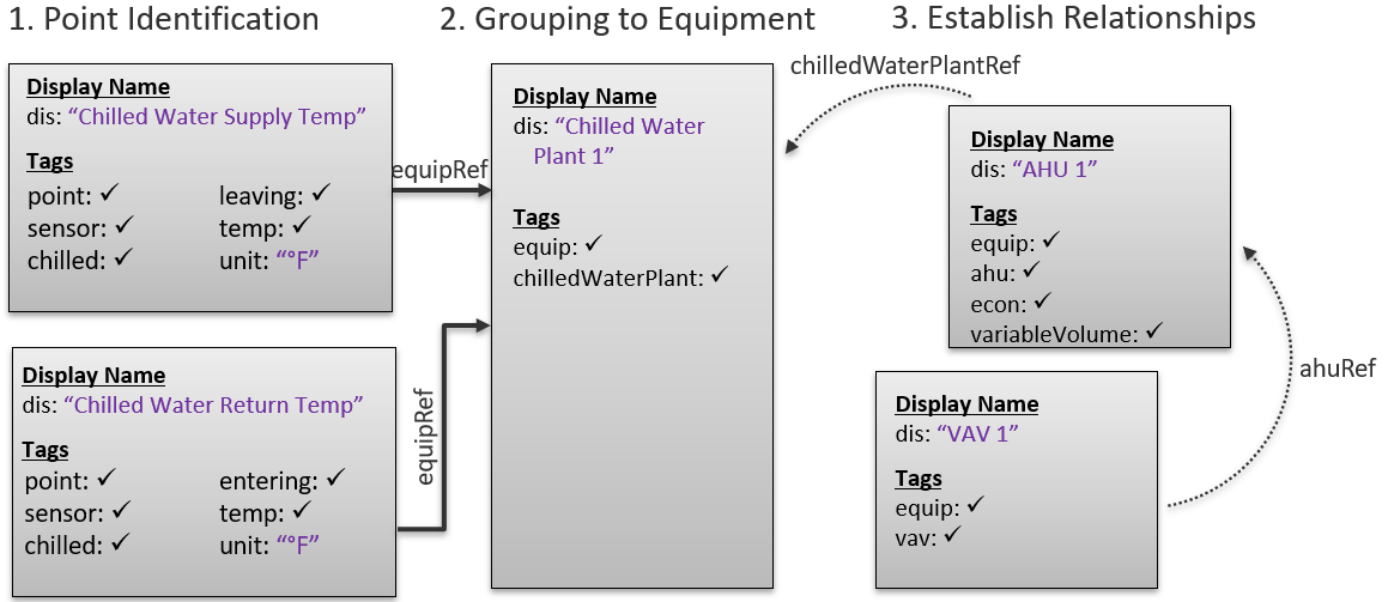

Figure 2 Conceptual three-step process for identifying Project Haystack tags

*1.2.1. Point Identification*

Point-type classification attempts to associate each BAS point to a specific function and define its specific characteristics. Point identification thus requires answering four key questions about each point entity [2]:

- What is the role of the point? In Haystack, a point may be classified as:
    - A sensor or input (`sensor` tag)
    - A command or output (`cmd` tag)
    - A setpoint, soft point, or calculated value (`sp` tag)
- What quantity is associated with the point? Typically, this will be a physical quantity such as temperature (`temp`), `pressure`, `flow`, or position (*e.g.*, `valve` position). Sometimes, it may be a control status (*e.g.*, `occ` for occupancy) or a network quantity, such as a data transfer rate.
- What substance or item is associated with the point? For example, a `sensor` might measure `air`, `water`, or `steam`. A `cmd` point might control a `damper` or `valve`.
- Where is the point located? Haystack tags such as `entering`, `leaving`, `discharge`, `return`, `exhaust`, and `outside` describe the location of the point (or its associated substance) within a system or piece of equipment.

The objective in point identification is to apply a set of descriptors (typically, marker tags) that uniquely distinguishes the point from all related points on the same piece of equipment.

*1.2.2. Equipment Identification*

In Haystack, points reside within an equipment entity (`equip` tag), such as an air handling unit (`ahu`), `chiller`, or electricity (`elec`) `meter`. Equipment items often have a standard or typical set of tags. For example, `ahu` equipment often have `outside`, `return`, `mixed`, and `discharge air temp sensor`s.

Equipment identification requires grouping associated points and using the point characteristics to infer the parent `equip`. Once a piece of equipment is identified, that information may be further leveraged to improve point classification, e.g., by identifying gaps in the point list and matching unmapped points by inference or process of elimination. Establishing the relationship between a point and its parent piece of equipment enables the application of an `equipRef` tag to the point, the value of which is the unique identifier of the parent piece of equipment.

### 1.2.3. System Identification

Individual pieces of equipment are typically organized into systems, e.g., a central `ahu` coupled to many VAV (`vav`) boxes via ductwork. System identification requires identifying these relationships and classifying them by type. For example, in Haystack a `vav` box is related to its central `ahu` via the `ahuRef` tag. An alternative expression of this process is visualizing equipment as nodes in a network (or graph) and the relationships as directed arcs between those nodes. The network represents the system of interest, which may be as large as the entire building.

The proposed UA address the first part of the problem definition, i.e., point identification and tagging.

## 3. DATA CONSIDERATIONS

### 3.1. Available Data Types

Raw BAS data can broadly be separated into two categories: (i) point data and (ii) time-series data. The point data may contain point name (likely not standardized); point address; device address, number, or name; point type (e.g., input, output, internal variable); and units. Raw point names typically contain useful information regarding the entity in question. Unfortunately, this information may be obscured by abbreviations or non-standard formatting conventions. For instance, Hong et al. point out that "SODA1R410B_ART", "SDF_SF1_R282_RMT", and "Zone Temp 2 RMI204" are all point names for air temperature sensors in different buildings [29]. In addition to point names, the other point tags may guide the point identification and equipment identification phases of tagging. For example, a point with units of *degrees Celsius* should have the tag `temp` and points located on the same panel are more likely to belong to the same system or piece of equipment.

Time-series data are generated by points throughout the building and are useful for clustering similar points. For instance, sensors measuring air temperature in different locations throughout the building are likely to report measurements that fall within the same general range of values and exhibit similar cyclic patterns. For this reason, time-series data are typically more consistent across buildings than raw point tags. However, extracting meaningful information from these data streams is difficult because they rarely capture complete or unique point type information. These data have been most effectively implemented for boosting performance of point tag assignment algorithms and for identifying interactions between points in a system.

### 3.2. Data Collection

EMIS used for analytics and supervisory control collect BAS data most often by polling, change-of-value subscription, or access to trend logs and/or a built-in historian (database of historical time-series data). In the case of polling or change-of-value, the connected EMIS collects data in real time for a period of days, weeks, or months in order to accumulate enough historical data for use in automated point mapping. If trend logs or a historian are available, access to previously collected historical data may speed this process. Ideally, a sufficiently long time period of data collection can be used to account for any cyclical trends (e.g., day/night, weekdays/weekends and seasonal changes) that may impact the data. In practice, waiting for enough data to account for seasonal trends lessens the value of the automation process. For this reason, in this work we use three weeks of data as a compromise that limits data collection time while still accounting for weekly cycles.

In addition to the time horizon of the data collection, the building type factors into the process. The algorithm development process for automated point mapping leverages data collected from several categories of buildings (e.g., retail, office, warehouse). Buildings are broken out categorically because different types of buildings are likely to have different distributions of sensors, actuators, etc. and may exhibit differences in variability in the point distributions from building to building—e.g., HVAC equipment and setup in retail buildings may tend to be relatively consistent, while office buildings may vary substantially.

### 3.3. Approaches for Tagging and impact of data availability

The choice of algorithms and the degree of effectiveness of the point tagging automation solution, to a certain extent, depend on temporal and spatial contexts in which the solution is being deployed. We present three likely scenarios that can be the starting point of the automation process. We also note building and HVAC type considerations that may assist in achieving tagging objectives. These approaches are discussed in the Appendix.

## 4. METHODS EMPLOYED

The proposed UA leverages as much information as possible in order to generate relevant metadata. Machine learning methods have proven to be powerful tools in data-driven clustering and classification problems. These methods uncover complex relationships in data to either organize the data in a meaningful way (unsupervised learning) or to learn mappings from inputs to outputs (supervised learning) [38]. Typically, ML methods leverage large amounts of data to uncover these relationships; however, the cases examined here are relatively data-sparse (compared to classical ML problems) and have a large output (tag) space. For this reason, we also employ domain knowledge into the unified architecture in the context of rule-based relationships. This section covers the background of the various rule-based, unsupervised, and supervised methods used in the unified architecture.

*4.1. Rule-based*

For certain commonly applied tags, there exist standard mutually exclusive relationships. This mutual-exclusivity can be exploited by embedding this information through rule-based programming in the framework. Some examples of such tags include the `sensor`/`sp` (or setpoint)/`cmd` (or command) tags as well as the `heat`/`cool` tags. A given point can only have one (or none) of these tags, but not multiple. We leverage this knowledge such that the application of one of these tags to a point precludes the possibility of applying the remaining tags.

Version 3 of the Project Haystack standard (the present version at the time of this writing) does not provide mutually exclusive tag sets in a machine-readable format. Therefore, we formulated machine-readable rules based on the human-readable logic described in the Project Haystack standard. We note that future versions of the Project Haystack standard are expected to include machine-readable content for mutually exclusive tags and similar tagging restrictions.

*4.2. Unsupervised clustering using tag names*

Raw point names can contain valuable insights in the metadata generation process. While naming and syntax conventions often vary significantly from building to building, point name data is typically more consistent within any individual building. This assumption underlies many of the rule-based approaches discussed in Section 1.2. We leverage this assumption here by performing unsupervised clustering on these strings to uncover useful structure within the data.

Mathematical clustering of points based on raw point names requires vectorization of the point name string. One common approach for text parsing is the *bag-of-words* vectorization method [39]. This process begins by tokenizing the point names into words based on various delimiters (e.g., spaces, underscores, etc.). The building's vocabulary is then constructed from all of the unique words identified, and each point is vectorized according to the frequency of each word in the point name. The bag-of-words approach suffers from several issues in the application of the methods to raw point name data in buildings. First, the approach is an orderless representation of the string. Typically, point names within a given building exhibit consistent ordering structures (e.g., building_floor_room_device); however, this information may not be available to the algorithm a priori. Furthermore, this method does not recognize abbreviations, acronyms, or misspellings of words, which are quite common in the raw point name data being clustered. Lastly, this approach is dependent on how words are delimited, which can vary from building to building.

A more robust approach to text vectorization employs *k*-mers, which were originally proposed as a method for comparing long strings of DNA [40]. *k*-mers have been used previously in metadata generation to effectively compare the complex, highly irregular point names within buildings [29]. The use of *k*-mers in this work differs in terms of how the clustered data is incorporated into the unified architecture. Given a string of length *L*, its *k*-mer decomposition is the collection of every substring of length *k*. Therefore, there is an inverse relationship between the number of *k*-mers exhibited by a word and the chosen value for *k*.

Traditional *k*-mer analysis is similar to the bag-of-word method where a vocabulary is constructed for the building based on the collection of unique *k*-mers identified among all of the point names. Each name is then vectorized based on the frequency of each *k*-mer in the string. However, this does not overcome issues related to ordering. We propose a new method for measuring *k*-mer similarity that better preserves ordering. The raw point name data is converted into a list of *k*-mers with the order maintained. We compute the similarity between the $i^{th}$ and $j^{th}$ *k*-mers from each pair of points,

$$s_{ij} = \begin{cases} 1 - \dfrac{|i-j|}{\max(N,M)} & \text{if } k\text{-mers match,} \\ 0 & \text{otherwise,} \end{cases} \quad (1)$$

where *N* and *M* are the number of *k*-mers in the two points names. Notice that this similarity metric rewards proximity of the *k*-mers within the strings; the metric is 1 if and only if the two points contain matching *k*-mers in the same position. This similarity decays to 0 as the positions of the matching *k*-mers deviate from each other. The total similarity of the two points names is simply the sum of the individual *k*-mer similarities,

$$s = \sum_{i=1}^{N} \sum_{j=1}^{M} s_{ij}. \quad (2)$$

Once the strings have been vectorized, an agglomerative hierarchical clustering approach is used to cluster the points [41]. This approach is chosen for two main reason: (i) it is simple to understand and implement and (ii) it is agnostic to the number of clusters in the data. This hierarchical clustering method initializes each data point as its own cluster and then iteratively combines the most similar clusters. Cluster similarity is measured according to the mean similarity metric in Eq. 2 between all points in each cluster.

The iterative procedure continues until the maximum cluster similarity dips below a given threshold. Figure 3 shows a similarity matrix for all of the points in a given building before and after clustering.

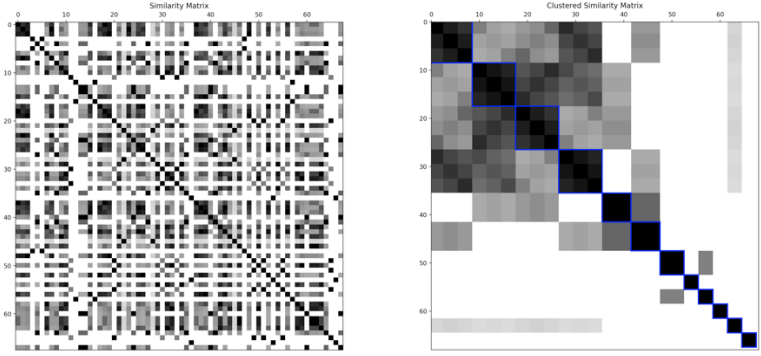

**Figure 3 Unclustered and clustered similarity matrices**

### 4.3. Supervised labeling using time-series data

Supervised learning models the relationship between input **x** and output data $y$ [42]. In the context of the metadata generation problem, the input data are the encoded time-series data streams from each point, and the output data are the applied Haystack tags. The UA described in Section 5 employs a two-step supervised learning approach that combines the output from a random forest algorithm and individualized support vector machine algorithms. In the next two sections, we briefly cover the background of these two methods.

#### 4.3.1. Random forest

Random forest models employ an ensemble of decision trees, which generate a graph-based model for classification. Nodes within this graphical structure examine various features of the input data and direct the flow of decision making until an output is achieved. Training a decision tree determines how the data is examined at each node in the graph. For instance, if a point is registering data that has mean values greater than 100, then a node in the decision tree may remove the possibility of the point measuring any percentage-based metric (e.g., relative humidity). Such models are powerful tools for classification problems, such as whether specific tags should or should not be applied to a given point.

The main issue with decision trees is that they tend to overfit to training data. Random forests address this by averaging across multiple decision trees that are trained on different subsets of the training data. This relatively simple fix has been found to result in significantly improved performance and robustness.

#### 4.3.2. Support vector machine

Support vector machines (SVM) are popular classification tools that divide the input domain into different regions based on the various output categories [43]. A linear SVM finds the optimal hyperplane that divides the data into two categories. That is, it attempts to fit

$$y = \boldsymbol{\theta}^\top \mathbf{x} + \theta_0, \tag{3}$$

while minimizing the magnitude of the parameters $\|\tilde{\boldsymbol{\theta}}\|$ where $\tilde{\boldsymbol{\theta}} = [\boldsymbol{\theta}^\top \quad \theta_0]^\top$. In reality, a perfectly discriminating hyperplane likely does not exist. Kernel SVM enables more flexibility by employing a nonlinear kernel mapping of the input space,

$$y = \sum_{i=1}^{N} \theta_i k(\mathbf{x}_i, \mathbf{x}) + \theta_0, \tag{4}$$

where $k(\cdot,\cdot)$ is a kernel function, such as the squared exponential kernel $k(\mathbf{x}_i, \mathbf{x}_j) = \exp\left(-\|\mathbf{x}_i - \mathbf{x}_j\|_2^2\right)$.

### 4.4. Rule-based versus machine-learning based algorithms - comparison in the context of this tagging problem

As described in Section 1.2, both rule-based and machine/statistical-learning-based methodologies have been utilized in various works in the literature with some degree of success. Table 1 summarizes the strengths and weaknesses of both these approaches in relation to the BAS metadata tagging problem. Note that the strengths of rule-based methods complement the weaknesses posed by machine-learning methods and vice-versa. This makes a strong case for leveraging both approaches in tandem to solve the problem. The next section presents the proposed Unified Architecture which employs both the approaches in the various decision stages.

| | Strengths | Weaknesses |
|---|---|---|
| Rule-based | • Are completely predictable (i.e., they behave exactly how humans have instructed them to)<br>• Provide the opportunity to embed human expertise in the program workflows | • Reflect biases or mistakes introduced by human error<br>• Infeasible to develop an all-encompassing rule-based algorithm that can cover all possible tagging combinations |
| Machine-learning based | • Capable of drawing inferences based on the training dataset (i.e., do not rely on human expertise)<br>• Efficient at finding patterns hidden underneath multi-dimensional dataset | • New tags not seen previously in training data will not be correctly applied<br>• Can be difficult to directly embed human knowledge in the process of drawing the inference |

Table 1 – Strengths and weaknesses of rule-based and ML-based tagging methods

5. PROPOSED UNIFIED ARCHITECTURAL FRAMEWORK

As explained in the previous section, both rule-based and machine-learning-based algorithms can contribute to the problem's solution, but neither is individually a "silver bullet." Automatically applying tags to the points in complex BASs is a challenging problem that cannot be effectively addressed at a practical scale by employing only one methodology, either rule-based or machine-learning based. Instead, the integration of both approaches with a cooperative decision-making paradigm yields the best results. Therefore, in this section, we propose a Unified Architecture (UA) framework (Figure 4) that synergistically combines multiple machine learning algorithms with a rule-based tagging scheme to infer correct tags by leveraging both time-series data and raw point name data. In the UA, the output from one decision block informs a subsequent decision block and the final decision of applying a specific tag to a specific point is therefore better-informed. This framework also provides the opportunity for leveraging human knowledge to inform rule-based tagging in the data preprocessing step to further improve the accuracy of the tags applied by the UA. To foster the success of the unified architecture in a real-time deployment environment, and increase its efficacy over time, a workflow for leveraging UA to build a long-term intelligent analytics platform for point tagging is presented as a part of data post-processing step.

The proposed framework consists of three main parts:

- Data preprocessing and model parameter inputs
- Unified architecture
- Data postprocessing

The following three sections explain these three components of the UA framework in detail.

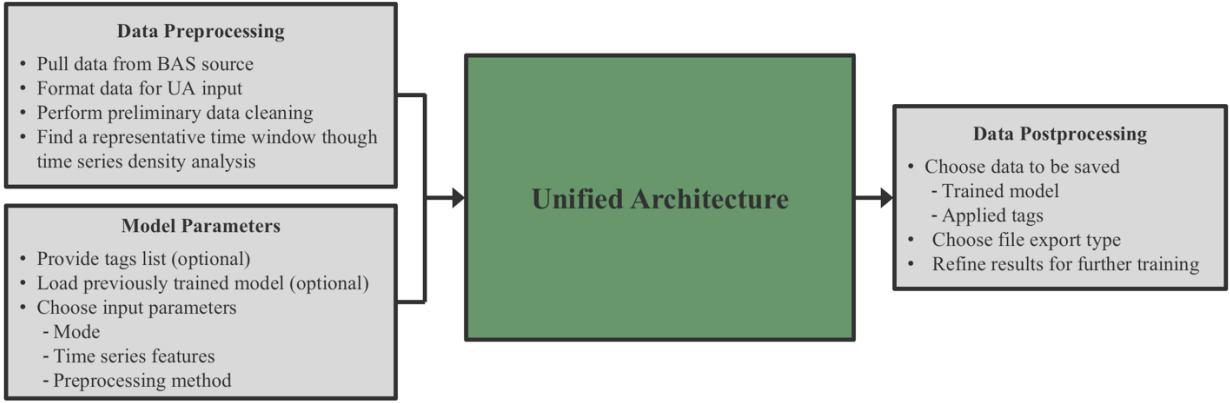

Figure 4 Overview of the Unified Architecture (UA) approach

*5.1. Data Preprocessing and Model Parameters*

For the points to be classified, the algorithm collects raw point names and associated time-series data. Time-series data is preprocessed for the supervised learning methods as follows. First, we find a representative time window where the building is generating data at a sufficiently high rate with all of the points reporting values without lapses in the interval data. This is important in new buildings where new points may be brought online slowly and may only be registering data sparsely at first. For this work, we find a three-week window of time-series data in order to capture daily and weekly trends that may be useful in point identification. Next, we compute various statistics, or features, of the time-series data and its gradient (e.g., the mean, median, min/max, variance, etc.) for each hour of the three-week window. We then compute the min/max, mean, and median for all of these intermediate features. Any point that is not registering at least five datapoints for 80% of the hours during the three weeks are filtered out. The vectorized time-series data is passed as inputs to the ML methods which predict the associated Haystack tags for each point.

To prepare the UA for training, the operator creates (or uses an existing) tag definition file that contains relevant tags and their relationships (Note: this step is optional since a default tags.csv is provided which contains all possible tags based on Haystack [10] standards). The UA's flexibility to ingest tag definition input provides an opportunity for embedding human knowledge about the possible tags for a specific building in the solution. At this time the operator also selects the transformation method used for converting input features into ML algorithm-compatible input: i) normalization, ii) standardization, or iii) min-max scaling. As an additional functionality useful in the real-time deployment context, the operator of the UA also selects one of the several modes of operation, including scratch training, supplemental/ updated training, or execution (application of tags); the purpose of these modes is explained in Section 5.3.

### 5.2. Unified Architecture

The core component of this framework is the unified architecture shown in Figure 5. The UA processes the input data through various stages and provides reliable inferences about the combinations of the tags which can be applied to a specific point. The UA is designed to promote cooperative decision-making between ML-based classifiers and fixed tagging rules. The architecture efficiently leverages all available information—both human knowledge and hidden patterns—to assign tags to points. The methodology of cooperative decision making and the information flow structure, employed by the UA is explained in detail in the remainder of this section.

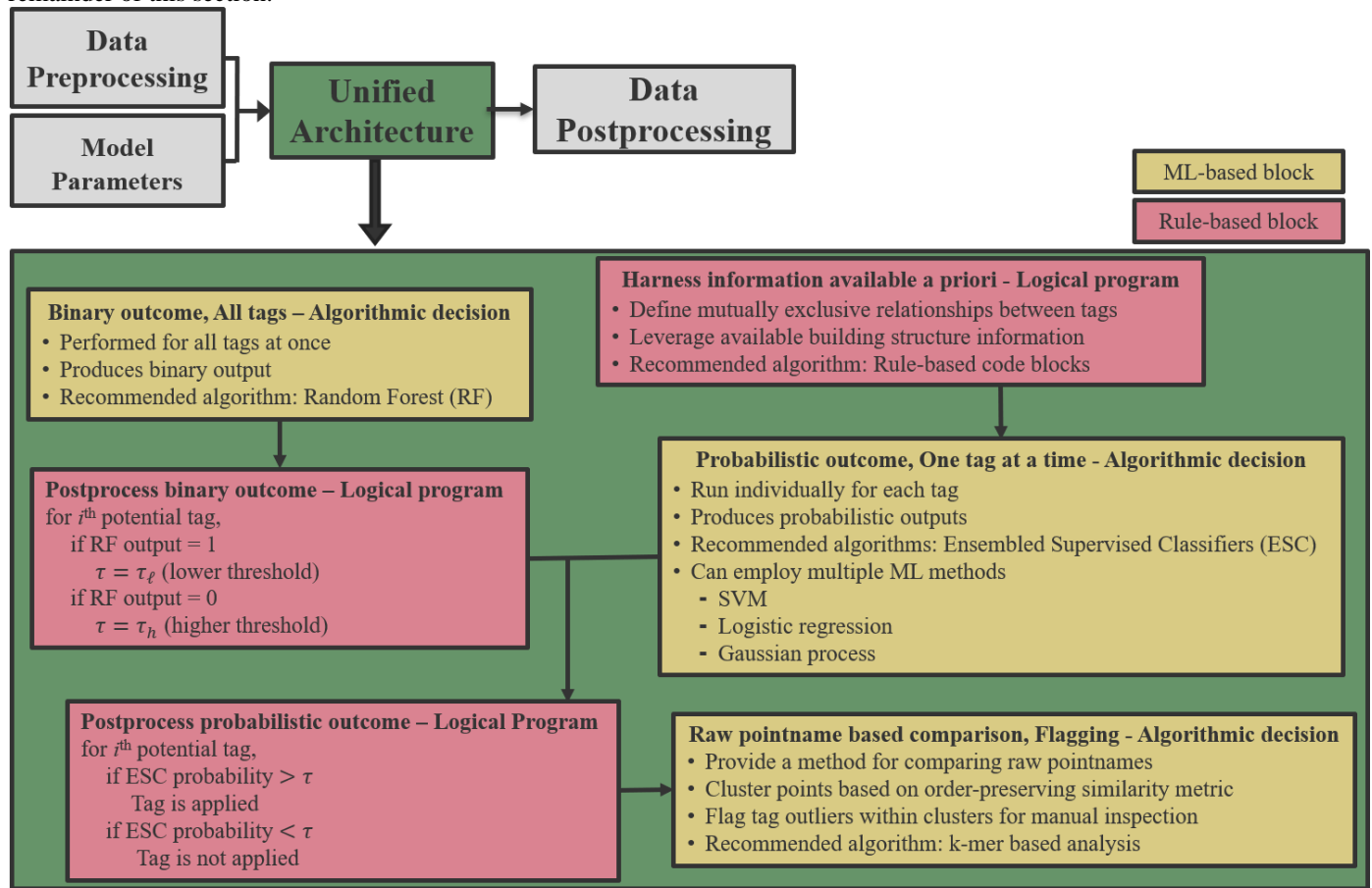

Figure 5 Callout of the UA to highlight the various individual components

The first branch of the UA is comprised of a machine learning block that uses a random forest algorithm to predict point tags. The random forest deterministically applies tags based on the time-series input data, as explained in Section 4.3. This algorithm considers the full space of all possible tags simultaneously (i.e., all-on-one). Based on the output of this model, a tag application threshold (denoted by $\tau$) is determined for each potential tag. This threshold is set lower if the random forest algorithm positively predicts a given tag and higher if it negatively predicts a given tag. The threshold values $\tau_\ell$ and $\tau_h$ are hyper-parameters that can be tuned based on human judgment of the performance of the UA for specific buildings.

The second branch of the UA combines a rule-based block and a machine learning block. The rule-based block leverages the mutually exclusive relationships provided in the tag definition file to enable the supervised learning classifiers to selectively apply these kinds of tags. This rule-based block, therefore, reduces the decision space for the ML algorithms by constraining the number of possible tags that can be applied to a single point. This information is passed to the ensembled supervised classifiers (ESC) block, which predicts tag individually (i.e., one-on-one) and iterates over all possible tags for each point to provide a probabilistic output. Although the block is set up to use any of a variety of user-specified supervised classification ML algorithms (e.g., SVM, logistic regression, and Gaussian processes), we determined empirically that SVM yields the most robust results for the data sets examined in this work. For each set of mutually exclusive tags, the ESC block is trained to assign probabilities for each tag such that the total probability never exceeds 1. The output from ESC is processed according to the rule-based filter that is set by the output from the random forest. This filter determines whether a given tag should be applied to a point by comparing the established threshold $\tau$ to the probabilistic output from the ESC. If the ESC probability is above the selected threshold, then the tag is applied. If it is below the threshold, then the tag is not applied. If we consider the random forest output as the baseline for whether or not a tag is applied, then $\tau_l$ requires the ESC probability be very low (i.e. the ESC is very confident that a given tag should not be applied) in order to remove the tag. Conversely, $\tau_h$ requires a higher confidence from the ESC in order to apply the tag.

After the tags have been applied, the UA employs a postprocessing. machine learning block that performs the unsupervised clustering approach based on the *k*-mer analysis described in Section 4.2. This block attempts to provide some measure of confidence on the applied tags and to flag those that could be wrong. To do so, the UA first clusters points names based on *k*-mer analysis of raw point names, then compares the applied tags of points within each cluster. The UA computes and applies outlier scores that are then used to identify points in the cluster which have markedly different tags applied. This process assumes that points with high similarity in the raw point names will have highly similar tags, and therefore flags outliers for manual inspection. This unsupervised learning block, therefore, provides an opportunity for leveraging operator knowledge most efficiently to manually assess only the tags applied to the points flagged by the *k*-mers block, for enhancing the accuracy of the overall result.

### 5.3. *Framework to enhance the effectiveness of UA through post-processing over time*

Once the tags are applied by the UA for a building, the postprocessing components of the framework are designed to facilitate long-term improvements in performance by harnessing a sophisticated data sampling and model retraining routine. As explained in Section 5.1, the UA has multiple operational modes that support includes training new models, loading and updating previously saved models, and performance testing. Figure 6 shows the functional relationships between these modes of operation, models, and data. The feedback arrows indicate an on-going data accumulation and curation routine that has the potential to contribute significantly to boosting the efficacy of the framework over the long term.

The operational modes of the UA are organized into i) a scratch training mode that trains new models from a full training dataset ii) a supplemental or updated training mode which loads previously trained models and updates them using limited training on new data, and iii) an execution mode which uses trained models to apply tags to new data sets. Of these modes, scratch training is the slowest, and applying tags is the fastest. Retraining is typically faster than scratch training but potentially may produce a less accurate model than the full training process. Therefore, we recommend occasional scratch training as the training dataset grows supplemented by retraining in the interim.

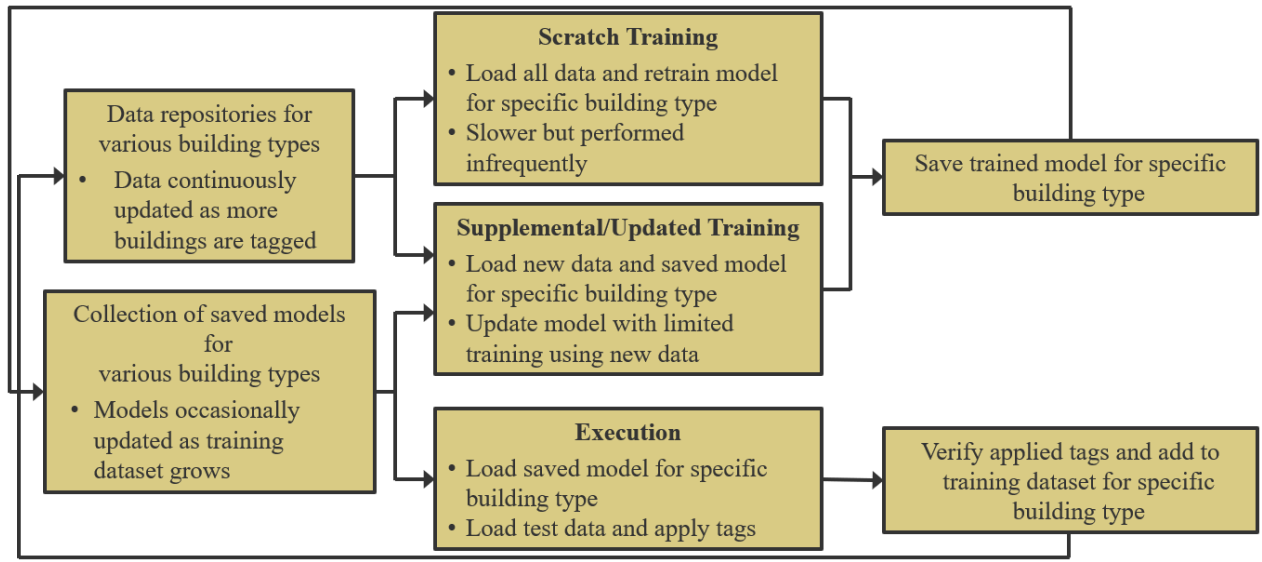

**Figure 6 Functional relationships between modes of model operations**

6. RESULTS AND DISCUSSION

We have implemented the proposed unified architecture in Python using the scikit-learn library. The target values (i.e., tags for each point in the dataset) for the training and the testing data samples were generated by manual application of Haystack tags to well-documented ground truth data sets. Two data sets were used to test the UA: i) commercial retail buildings and ii) National Renewable Energy Laboratory (NREL) campus office buildings. The next two sections present the results obtained for the two case studies, followed by a subsection discussing the limitations of the proposed Unified Architecture.

*6.1. Test Case I - Commercial retail building*

The first test case for the UA examines three small commercial retail buildings. The three buildings contain similar HVAC systems and have previously been manually tagged The main equipment in these buildings are multiple rooftop air handling units. During data preprocessing, we filter out points with insufficient data to tag automatically. After filtering, each of the three buildings contains approximately 40-60 unique points to be tagged. For the study, we first trained the UA on two of the buildings and tested the performance on the third, repeating this process for each of the three buildings. The results of this case study are as follows. We first examine the tagging results for each of the three buildings. Table 2 compares the F1 scores for component standalone tagging methods to the UA approach which combines multiple algorithms. We can see that by combining outputs from multiple algorithms, we obtain better tagging results than from a single algorithm approach. Figure 7 shows the proportion of true positives, false positives, and false negatives relative to the total number of tags in each building. In all three buildings, the UA was able to correctly apply between 85-90% of the tags while incurring approximately a 10% rate for false positives and false negatives. (We do not include the results for true negatives because they massively outweigh the total number of correct tags for the building.) Figure 7 also includes the F1 score for the UA across the three buildings. In each case, the F1 score is between 0.87-0.90. Thus, the algorithm was able to effectively learn the characteristics of the time-series training data and use it to apply tags to new buildings.

|  | Building A | Building B | Building C |
|---|---|---|---|
| Random Forest | 0.7098 | 0.7114 | 0.6868 |
| Support Vector Machine | 0.8325 | 0.8375 | 0.8486 |
| Unified Architecture | 0.8994 | 0.9038 | 0.8782 |

**Table 2 – Comparison of a single algorithm tagging results versus the Unified Architecture approach**

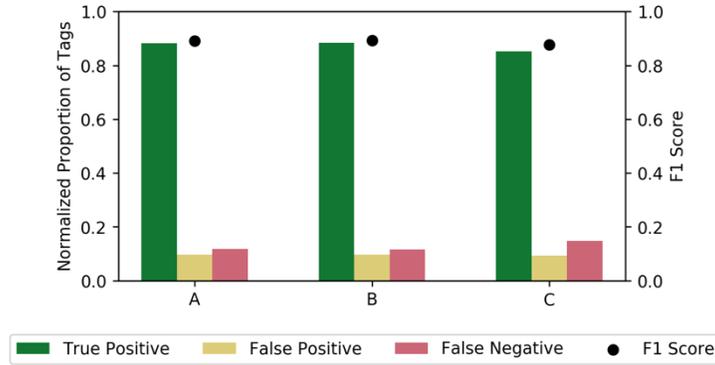

**Figure 7 Performance of the UA on commercial retail buildings**

Next, we examine the performance of the UA with regards to the specific tags. Figure 8 shows the individual F1 scores for the UA for various tags that are present in the commercial retail buildings. Additionally, the plot shows the total number of times each tag appears across the three buildings. There is a wide range of performance across the various tags. For example, the algorithm was able to effectively identify setpoints (denoted `sp`) and command points (denoted `cmd`), but struggled to correctly tag sensors. Additionally, we see a mild correlation between the frequency of tags and the ability of the UA to correctly apply them. Specifically, any tags that appeared at least 40 times had an F1 score of at least 0.9. However, several tags with low counts were still able to be correctly identified with high accuracies, such as the `min`/`max` tags as well as the `humidity` tag. This could be due to the associated points exhibiting a highly unique time-series fingerprint.

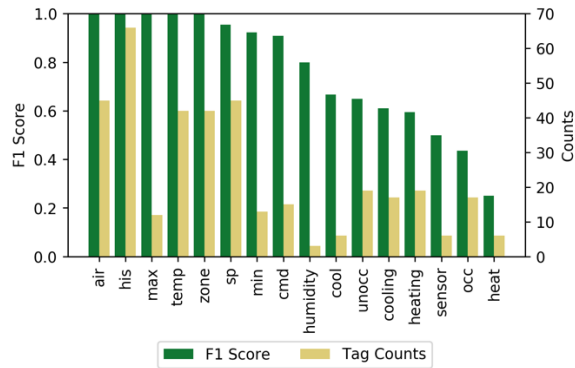

**Figure 8 Performance of the UA for specific tags in the commercial retail buildings**

Lastly, we briefly examine the outlier detection capability of the UA. Recall that this feature clusters points using the order-preserving *k*-mer analysis presented in Section 4.2 and then compares the applied tags across various points in the cluster. The assumption is that points with similar raw point names within a building should have similar tags applied. As discussed earlier, this assumption will not likely be valid in all cases, which is why this method is only used to flag potential errors for human inspection. Table 3 contains an example of the outlier detection working well. It contains three points from one of the retail buildings that were clustered based on their raw point names. The point with the highest outlier score within the cluster (SP_UNOCC_COOL) did indeed have several tags missing (the `cooling` and `unocc` tags). Additionally, notice that the other points in the cluster have nonzero outlier scores despite being correctly tagged. This is due to differences in the tags to the tags applied to SP_UNOCC_COOL as well as the minor difference in tags between the two points. This highlights that the assumption underlying this metric is not applicable in all the cases but that it does provide a useful tool for indicating which points may require human inspection.

| Raw Point Name | SP_UNOCC_COOL | SP_UNOCC_COOL_MAX | SP_UNOCC_COOL_MIN |
|---|---|---|---|
| Outlier Score | 0.22 | 0.17 | 0.17 |
| True Positives | air<br>his<br>point<br>sp<br>temp<br>zone | air<br>cooling<br>his<br>max<br>point<br>sp | air<br>cooling<br>his<br>min<br>point<br>sp |

|  |  |  | `temp unocc zone` | `temp unocc zone` |
|---|---|---|---|---|
| False Positives | None | None | None | None |
| False Negatives | `cooling unocc` | None | None | None |

Table 3 – Example of outlier detection identifying errors in the applied tags

*6.2. Test Case II – NREL campus*

Next, we examine data from the NREL campus. This data comes from the NREL Energy Systems Integration Facility (ESIF) and contains 352 points. These points are primarily associated with the larger air systems in the building, including multiple AHUs, makeup air units (or MAUs), and exhaust fans. For this study, we divide the data into five approximately equal subsets to study the performance of the UA. Similar to the previous study, in each case we train the UA on four subsets and test it on the fifth.

The results of applying the UA to the NREL data are shown in Figure 9. As previously, the figure shows the proportion of true positives, false positives, and false negatives normalized by the total number of tags in each subset. Overall, the UA did not perform as well on the NREL data as it did on the commercial retail data. In particular, the algorithm produced more false negatives, that is, it did not apply tags that it should have applied. Despite this, the UA was still able to correctly apply 70-75% of the tags to the test data and obtained an F1 score of approximately 0.8 in all test cases.

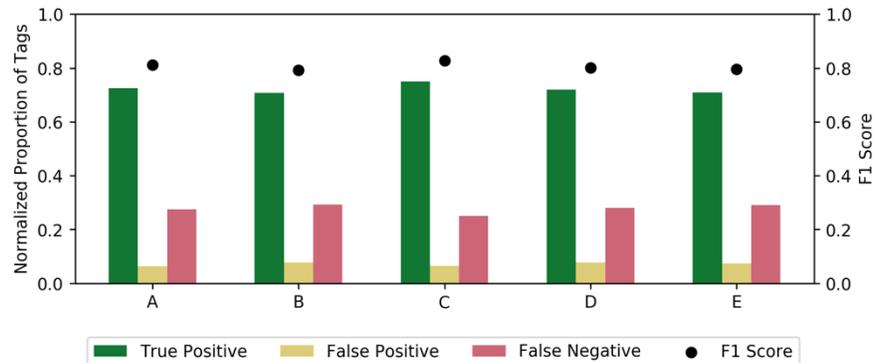

**Figure 9 Performance of the UA on NREL buildings**

*6.3. Challenges and limitations*

A central challenge to any machine learning framework is effectively dealing with data quality concerns. Several potential sources of error in the training data could originate from malfunctioning devices or mislabeled points. The proposed UA attempts to mitigate these concerns by including a time-series density analysis in the data preprocessing to ensure time-series data is being generated by a sufficient number of points. The *k*-mer postprocessing can also help identify issues related to data quality by flagging outlier points that appear to have a mismatch between the raw point name and the applied tags. While these features may prevent against minor data quality concerns, more systematic data errors would be difficult for the proposed UA to overcome. Identifying such scenarios is one possible research direction towards enhancing these types of automated point tagging systems.

The purpose of the BAS tagging process is to identify all of the datapoints contextually and generate relevant metadata associated with them. The proposed UA solves only the first piece of the puzzle, i.e., point tagging. To address the problem of automated BAS tagging process in its entirety, the remaining two pieces – i) equipment identification, and ii) system identification need to be addressed. Also, currently, the UA is designed to maintain different machine learning models for different types of buildings, this is because using a single model for all types of buildings can potentially deteriorate the performance of the models. Although the lack of a single model to address all types of buildings is a practical limitation, if the models are effectively managed the use of multiple models can lead to better performance than a single model. In addition, the UA requires tuning hyperparameters for multiple ML models as well as several hyperparameters for the UA itself; further research and testing with larger data sets will inform the optimal selection of these hyperparameters.

7. CONCLUSIONS AND FUTURE DIRECTIONS

In this article, we present a Unified Architecture that combines human-generated rule-based logic with data-driven ML methods to apply Haystack tags to BAS points. By employing an approach that intertwines both the techniques, the algorithm leverages as much knowledge as possible in applying these tags, including both prior knowledge of tag relationships and raw point name and time-series data generated by the BAS. The UA algorithm shows high performance in tagging points in two test cases: commercial retail and office building settings. We also propose a framework to deploy in real-life settings and further enhance the performance of the proposed Unified Architecture with time by retraining models with newly accumulated data.

Haystack tags provide invaluable metadata that is used by a BAS to intelligently drive a building's energy usage. However, this metadata does not present the full picture of a building's infrastructure. Once points are identified via the tagging process, larger pieces of equipment and systems must be identified. This direction of inquiry will be characterized as a 'bottom-up' approach, where various equipment are identified with the information that is made available through point tags applied by the UA. This problem, which is left as future research directions, maybe approached using a similar hybrid methodology as the one presented here. That is, a unified algorithm that leverages rule-based knowledge (e.g., what points are present in a given building, what collections of points constitute various pieces of equipment, etc.) and data-driven techniques (e.g., analyzing time-series correlations to identify operational relationships, learning $k$-mers that identify equipment clusters, etc.). We anticipate that further improvements in Haystack and other BAS metadata standards will enable a more precise, better-automated application of rules within the UA to constrain the tagging output space.

Another promising future research direction is taking a 'top-down' approach where performing equipment and system identification before tagging individual points will help inform the point 'typing' process as opposed to being a sequential step that is executed after the points are tagged using the proposed UA. Additionally, building vintage (or most recent retrofit vintage) and statistical analysis of implemented mechanical systems by building type can also be factored into this analysis as another layer of *apriori* knowledge being fed to the UA. Similar to how Commercial Buildings Energy Consumption Survey (CBECs) has informed the development of the systems and equipment for the prototype buildings, the data about when the building was built and the type of building has the potential to provide significant information on typical systems for optimizing the decision-space for the ML algorithms, to help establish the relationships of the points of the equipment in the buildings, and the typical points associated with equipment at different points in time.

ACKNOWLEDGMENT

This work was authored by the National Renewable Energy Laboratory (NREL), operated by Alliance for Sustainable Energy, LLC, for the U.S. Department of Energy (DOE) under Contract No. No. DE-AC36-08GO28308 in collaboration with Brainbox AI. This work was supported by the cooperative research and development agreement CRD-18-00767. The authors wish to thank Cory Mosiman for providing useful suggestions to refine the manuscript. The views expressed in the article do not necessarily represent the views of the DOE or the U.S. Government. The U.S. Government retains and the publisher, by accepting the article for publication, acknowledges that the U.S. Government retains a nonexclusive, paid-up, irrevocable, worldwide license to publish or reproduce the published form of this work, or allow others to do so, for U.S. Government purposes.

APPENDIX

For real-life deployment of automated metadata tagging systems, the role of data availability as well as the nature of the collected data is crucial. In the following sub-sections, we describe various scenarios, that may arise in real-time, to which the proposed UA may be adapted.

### 9.1. Tagging a building from scratch

If the building in question (referred to as target building here on) is new, suggesting no time-series data availability, no tagged-points are available to recognize the pattern in the point names available in BAS. Also, if the building's control architecture does not have any similarity with an older and already tagged building (referred to as source building here on), then the scope of the automation solution is to tag a building from scratch, without any foundational information source available a priori. This scenario can be handled in a few different ways:

#### 9.1.1. Manual tagging with a subset of points

In this methodology, an expert examines the naming convention used in the target building and leverages knowledge of the building systems to then tag a small percentage of the overall points in the building, e.g. 20%. With this information in hand, semi-supervised learning approaches can be employed to tag the remained of the points. This could yield another ~60-65% of the points being correctly tagged through the automation process. The remaining ~15-20% points would then be tagged manually. Text parsing approaches can also be applied for this kind of scenario: the examples which are parsed by the expert, in the beginning, can be chosen in an intelligent way by clustering the points a priori.

#### 9.1.2. Accumulating time-series data – limited duration

This methodology leans toward working with time-series data of the target building itself. The data from the first few weeks of operation can be buffered to conduct unsupervised learning on the accumulated time-series data. Due to the lack of any other information about the syntactical pattern employed for naming the BAS points, this method is likely not going to produce high accuracies of the automated tagging.

*9.2. Tagging a building similar to an already tagged building*

If the control architecture of the target building is similar to the architecture of the source building, then it is possible to focus the scope of the automation solution toward transfer learning approaches. The algorithms can be trained on the source building's time-series data. The same models can be deployed for the target building with a minimal amount of time-series data accumulated over the first few days of operation of the building.

*9.3. Tagging a building with a substantial amount of time-series data availability*

If the target building has been accumulating data for some period, this opens the opportunity to leverage the accumulated time-series data of the building to train. The problem can then be modeled as a time-series classification problem. There are no prior tags available to train the model on, but the patterns in the time-series data (measurements, control signal, setpoints) can be captured to classify the points, as the first step in the process. The same time-series data can then be employed for discovering the spatial relationships between sensors and building spaces, geared toward objective two (equipment identification) and three (spatial relationships within the equipment) fulfillment.

*9.4. Type of commercial building*

The model and the associated algorithm can be targeted for specific types of buildings (for example retail, office, warehouse, hospital, university campus). This will enable the automation solution to leverage transfer learning approaches more effectively. The type of building classification is mainly useful in the second and third objectives of the automation process: equipment identification and system identification. The point classification model, in general, is largely independent of the type of building.

*9.5. Type of HVAC Technology*

Targeting the algorithm based on the type of HVAC technology used for the specific building is potentially another effective way of leveraging the contextual information to tailor the solution for the second objective, the equipment classification. Various types of HVAC technologies are used in commercial buildings including 1) Variable Air Volume (VAV), 2) Constant Air Volume (CAV), 3) Variable Refrigerant Flow (VRF), 4) Chilled beams (active and passive), 5) Heat Pump, 6) Fan Coils (FCU) and Blower Coils (BCU), etc. If the set of HVAC technologies in use is known, then the tags to be applied may be tailored to match the technology space.

References

[1] The American Society of Heating, Refrigerating and Air-Conditioning Engineers, "Standard 135-2016 -- BACnet-A Data Communication Protocol for Building Automation and Control Networks," 2016. [Online]. Available: https://www.techstreet.com/ashrae/standards/ashrae-135-2016?ashrae_auth_token=&product_id=1918140#jumps. [Accessed 16 July 2020].

[2] Echelon Corporation, "Introduction to the LonWorks Platform." 2009. https://www.echelon.com/assets/blt893a8b319e8ec8c7/078-0183-01B_Intro_to_LonWorks_Rev_2.pdf. [Accessed 16 July 2020].

[3] Tridium, Inc., "Niagara AX Networking and IT Guide," 2006. https://docplayer.net/17972259-Technical-document-niagara-ax-networking-and-it-guide-october-9-2006.html. [Accessed 16 July 2020].

[4] KNX Norway, "A brief introduction to KNX," National KNX Norway, [Online]. Available: https://www.knx.org/knx-en/for-professionals/What-is-KNX/A-brief-introduction/index.php. [Accessed 16 July 2020].

[5] Lawrence Berkeley National Laboratory, "A primer on organizational use of energy management and information systems," United States Department of Energy, 2015. https://betterbuildingssolutioncenter.energy.gov/resources/a-primer-organizational-use-energy-management-and-information-systems-emis. [Accessed 16 July 2020].

[6] M. Neukomm, V. Nubbe and R. Fares, "Grid-interactive efficient buildings," United States Department of Energy Office of Energy Efficiency & Renewable Energy, 2019. https://doi.org/10.2172/1508212

[7] D. Cutler, S. Frank, M. Slovensky, M. Sheppy and A. Petersen, "Creating an Energy Intelligent Campus: Data Integration Challenges and Solutions at a Large Research Campus," in *American Council for an Energy-Efficient Economy Summer Study on Energy Efficiency in Buildings*, Pacific Grove, CA, USA, p. 12, 2016. http://aceee.org/files/proceedings/2016/data/index.htm. [Accessed 16 July 2020].

[8] D. Hardin, C. Corbin, E. Stephan, S. Widergren and W. Wang, "Buildings interoperability landscape," United States Department of Energy Pacific Northwest National Laboratory, Technical Report PNNL-25124, 2015. https://doi.org/10.2172/1234792

[9] Energy Information Administration, "Commercial Buildings Energy Consumption Survey," 2012. [Online]. Available: https://www.eia.gov/consumption/commercial/data/2012/. [Accessed 16 July 2020].

[10] Project Haystack Corporation, "Project Haystack," [Online]. Available: https://project-haystack.org/. [Accessed 16 July 2020].

[11] Brick Development Team, "Brick - A uniform metadata schema for buildings," [Online]. Available: https://brickschema.org/. [Accessed 16 July 2020].

[12] J. Granderson and G. Lin, "Building energy information systems: synthesis of costs, savings, and best-practice uses," *Energy Efficiency*, vol. 9, no. 6, pp. 1369-1384, 2016. https://doi.org/10.1007/s12053-016-9428-9

[13] S. Frank, X. Jin, D. Studer and A. Farthing, "Assessing barriers and research challenges for automated fault detection and diagnosis technology for small commercial buildings in the United States," *Renewable and Sustainable Energy Reviews*, vol. 98, pp. 489-499, 2018. https://doi.org/10.1016/j.rser.2018.08.046

[14] A. Bhattacharya, D. E. Culler, J. Ortiz, D. Hong and K. Whitehouse, "Enabling portable building applications through automated metadata transformation," Electrical Engineering & Computer Sciences Department, University of California, Berkeley, Technical Report UCB/EECS-2014-159, 2014. http://www2.eecs.berkeley.edu/Pubs/TechRpts/2014/EECS-2014-159.html. [Accessed 16 July 2020].

[15] A. Schumann, J. Ploennigs and B. Gorman, "Towards automating the deployment of energy saving approaches in buildings," in *Proceedings of the 1st Association for Computing Machinery Conference on Embedded Systems for Energy-Efficient Buildings*, Memphis, TN, USA, 2014, pp. 164-167. https://doi.org/10.1145/2674061.2674081

[16] D. Hong, H. Wang and K. Whitehouse, "Clustering-based active learning on sensor type classification in buildings," in *Proceedings of the 24th Association for Computing Machinery International on Conference on Information and Knowledge Management*, Melbourne, Australia, 2015, pp.363-372. https://doi.org/10.1145/2806416.2806574

[17] P. Esling and C. Agon, "Time-series data mining," *Association for Computing Machinery Computing Surveys*, vol. 45, no. 1, p. 12, 2012. https://doi.org/10.1145/2379776.2379788

[18] D. Hong, J. Ortiz, K. Whitehouse and D. Culler, "Towards automatic spatial verification of sensor placement in buildings," in *Proceedings of the 5th Association for Computing Machinery Workshop on Embedded Systems For Energy-Efficient Buildings*, Roma, Italy, 2013, pp. 1-8. https://doi.org/10.1145/2528282.2528302

[19] N. E. Huang, Z. Shen, S. R. Long, M. C. Wu, H. H. Shih, Q. Zheng, N.-C. Yen, C. C. Tung and H. H. Liu, "The empirical mode decomposition and the Hilbert spectrum for nonlinear and non-stationary time series analysis," *Proceedings of the Royal Society of London. Series A: Mathematical, Physical and Engineering Sciences*, vol. 454, no. 1971, pp. 903-995, 1998. https://doi.org/10.1098/rspa.1998.0193

[20] M. Koc, B. Akinci and M. Berges, "Comparison of linear correlation and a statistical dependency measure for inferring spatial relation of temperature sensors in buildings," in *Proceedings of the 1st Association for Computing Machinery Conference on Embedded Systems for Energy-Efficient Buildings*, Memphis, TN, USA, 2014, pp. 152-155. https://doi.org/10.1145/2674061.2674075

[21] B. Akinci, M. Berges and A. G. Rivera, "Exploratory study towards streamlining the identification of sensor locations within a facility," in *Computing in Civil and Building Engineering*, Orlando, FL, USA, 2014, pp. 1820-1827. https://doi.org/10.1061/9780784413616.226

[22] J. Gao, J. Ploennigs and M. Berges, "A data-driven meta-data inference framework for building automation systems," in *Proceedings of the 2nd Association for Computing Machinery International Conference on Embedded Systems for Energy-Efficient Built Environments*, Seoul, South Korea, 2015, pp. 23-32. https://doi.org/10.1145/2821650.2821670

[23] J.-P. Calbimonte, O. Corcho, Z. Yan, H. Jeung and K. Aberer, "Deriving semantic sensor metadata from raw measurements," in *Proceedings of the 5th International Conference on Semantic Sensor Networks*, Boston, MA, USA, 2012, pp. 33-48. https://dl.acm.org/doi/10.5555/2887689.2887692

[24] E. Holmegaard and M. B. Kjærgaard, "Mining building metadata by data stream comparison," in *2016 Institute of Electrical and Electronics Engineers Conference on Technologies for Sustainability*, Phoenix, AZ, USA, 2016, pp. 28-33. https://doi.org/10.1109/SusTech.2016.7897138

[25] J. Koh, D. Hong, R. Gupta, K. Whitehouse, H. Wang and Y. Agarwal, "Plaster: an integration, benchmark, and development framework for metadata normalization methods," in *Proceedings of the 5th Conference on Systems for Built Environments*, Shenzhen, China, 2018, pp. 1-10. https://doi.org/10.1145/3276774.3276794

[26] J. Koh, B. Balaji, D. Sengupta, J. McAuley, R. Gupta and Y. Agarwal, "Scrabble: transferable semi-automated semantic metadata normalization using intermediate representation," in *Proceedings of the 5th Conference on Systems for Built Environments*, Shenzen, China, 2018, pp. 11-20. https://doi.org/10.1145/3276774.3276795

[27] A. A. Bhattacharya, D. Hong, D. Culler, J. Ortiz, K. Whitehouse and E. Wu, "Automated metadata construction to support portable building applications," in *Proceedings of the 2nd Association for Computing Machinery International Conference on Embedded Systems for Energy-Efficient Built Environments*, Seoul, South Korea, 2015, pp. 3-12. https://doi.org/10.1145/2821650.2821667

[28] B. Balaji, C. Verma, B. Narayanaswamy and Y. Agarwal, "Zodiac: organizing large deployment of sensors to create reusable applications for buildings," in *Proceedings of the 2nd Association for Computing Machinery International Conference on Embedded Systems for Energy-Efficient Built Environments*, Seoul, South Korea, 2015, pp. 13-22. https://doi.org/10.1145/2821650.2821674

[29] D. Hong, H. Wang, J. Ortiz and K. Whitehouse, "The building adapter: towards quickly applying building analytics at scale," in *Proceedings of the 2nd Association for Computing Machinery International Conference on Embedded Systems for Energy-Efficient Built Environments*, Seoul, South Korea, 2015, pp. 123-132. https://doi.org/10.1145/2821650.2821657

[30] L. Kidd, *Auto-tagging with machine learning* at Haystack Connect 2019, San Diego, CA, USA, 15 May 2019. [Online]. Available: https://www.haystackconnect.org/wp-content/uploads/2019/05/Auto-Tagging-with-Machine-Learning-Lucy-Kidd.pdf. [Accessed 30 August 2020].

[31] B. Butzin, F. Golatowski and D. Timmermann, "A survey on information modeling and ontologies in building automation," in *43rd Annual Conference of the Institute of Electrical and Electronics Engineers Industrial Electronics Society*, Beijing, China, 2017, pp. 8615-8621. https://doi.org/10.1109/IECON.2017.8217514

[32] J. Ploennigs, B. Hensel, H. Dibowski and K. Kabitzsch, "BASont - a modular, adaptive building automation system ontology," in *38th Annual Conference on Institute of Electrical and Electronics Engineers Industrial Electronics Society*, Montreal, Quebec, Canada, 2012, pp. 4827-4833. https://doi.org/10.1109/IECON.2012.6389583

[33] International Organization for Standardization, ISO/PAS 16739:2005 Industry Foundation Classes, Release 2x, Platform Specification (IFC2x Platform), International Organization for Standardization, 2005. https://www.iso.org/standard/38056.html. [Accessed 16 July 2020].

[34] F. T. H. den Hartog, L. M. Deniele and J. Roes, "Study on semantic assets for smart appliances interoperability: D-S1: First interim report," European Union, 2014. https://repository.tudelft.nl/view/tno/uuid%3A4b094d36-a2d1-4f2f-bef1-0034711e9d8c. [Accessed July 16 2020].

[35] B. Balaji, A. Bhattacharya, G. Fierro, J. Gao, J. Gluck, D. Hong, A. Johansen, J. Koh, J. Ploennigs, Y. Agarwal, M. Berges, D. Culler, R. Gupta, M. B. Kjærgaard, M. Srivastava and Whiteh, "Brick: Towards a unified metadata schema for buildings," in *Proceedings of the 3rd Association for Computing Machinery International Conference on Systems for Energy-Efficient Built Environments*, Palo Alto, CA, USA, 2016, pp. 41-50. https://doi.org/10.1145/2993422.2993577

[36] B. Balaji, A. Bhattacharya, G. Fierro, J. Gao, J. Gluck, D. Hong, A. Johansen, J. Koh, J. Ploennigs, Y. Agarwal, M. Berges, D. Culler, R. Gupta, M. B. Kjærgaard, M. Srivastava and Whiteh, "Brick: Metadata schema for portable smart building applications," *Applied Energy*, vol. 226, pp. 1273-1292, 2018. https://doi.org/10.1016/j.apenergy.2018.02.091

[37] O. Lassila and R. Swick, "Resource description framework (RDF) model and syntax specification," 1999. https://www.w3.org/TR/1999/REC-rdf-syntax-19990222/. [Accessed 16 July 2020].

[38] C. M. Bishop, Pattern Recognition and Machine Learning, New York: Springer, 2006. ISBN: 0387455280.

[39] Y. Zhang, R. Jin and Z.-H. Zhou, "Understanding bag-of-words model: a statistical framework," *International Journal of Machine Learning and Cybernetics*, vol. 1, no. 1-4, pp. 43-52, 2010. https://doi.org/10.1007/s13042-010-0001-0

[40] P. A. Pevzner, H. Tang and M. S. Waterman, "An Eulerian path approach to DNA fragment assembly," *Proceedings of the National Academy of Sciences*, vol. 98, no. 17, pp. 9748-9753, 2001. https://doi.org/10.1073/pnas.171285098

[41] L. Rokach and O. Maimon, "Clustering Methods," in Data Mining and Knowledge Discovery Handbook, Boston: Springer, 2005, pp. 321-352. ISBN: 9780387244358

[42] T. K. Ho, "Random decision forests," in *Proceedings of 3rd International Conference on Document Analysis and Recognition*, Montreal, Quebec, Canada, 1995, pp. 278-282. https://doi.org/10.1109/ICDAR.1995.598994

[43] V. N. Vapnik, The Nature of Statistical Learning Theory, New York: Springer, 2000. ISBN: 1475724403.